\documentstyle[12pt]{article}
\def\ie{{\em i.e.}}
\def\eg{{\em e.g.}}

\def\beq{\begin{equation}}
\def\eeq{\end{equation}}

\catcode`\@=11 

\def\lsim{\mathrel{\mathpalette\@versim<}}
\def\gsim{\mathrel{\mathpalette\@versim>}}
\def\@versim#1#2{\vcenter{\offinterlineskip
    \ialign{$\m@th#1\hfil##\hfil$\crcr#2\crcr\sim\crcr } }}
\def\etal{{\em et. al.}}
\def\JL{J. L. Lopez}
\def\DVN{D. V. Nanopoulos}
\def\AZ{A. Zichichi}

\def\t1{{\tilde 1}}

\def\GeV{\,{\rm GeV}}

\def\to{\rightarrow}

\def\NPB#1#2#3{Nucl. Phys. B {\bf#1} (19#2) #3}
\def\PLB#1#2#3{Phys. Lett. B {\bf#1} (19#2) #3}

\def\PRD#1#2#3{Phys. Rev. D {\bf#1} (19#2) #3}
\def\PRL#1#2#3{Phys. Rev. Lett. {\bf#1} (19#2) #3}
\def\PRT#1#2#3{Phys. Rep. {\bf#1} (19#2) #3}
\def\MODA#1#2#3{Mod. Phys. Lett. A {\bf#1} (19#2) #3}
\def\IJMP#1#2#3{Int. J. Mod. Phys. A {\bf#1} (19#2) #3}

\begin{document}
\begin{flushright}
\baselineskip=12pt
{CTP-TAMU-25/95}\\
{ACT-10/95}\\
{\tt hep-ph/9506217}\\
\end{flushright}

\begin{center}
\vglue 1cm
{\Huge\bf R$_{b}$ in supersymmetric models\\}
\vglue 2cm
{XU~WANG$^{1,2}$, JORGE L. LOPEZ$^{1,2}$, and D. V. NANOPOULOS$^{1,2,3}$\\}
\vglue 1cm
{\em $^{1}$Center for Theoretical Physics, Department of Physics, Texas A\&M
University\\}
{\em College Station, TX 77843--4242, USA\\}
{\em $^{2}$Astroparticle Physics Group, Houston Advanced Research Center
(HARC)\\}
{\em The Mitchell Campus, The Woodlands, TX 77381, USA\\}
{\em $^{3}$CERN Theory Division, 1211 Geneva 23, Switzerland\\}
\baselineskip=12pt
\end{center}

\vglue 1cm
\begin{abstract}
We compute the supersymmetric contribution to $R_{b}\equiv \Gamma (Z\to
b{\bar b})/\Gamma (Z\to {\rm hadrons})$ in a variety of supersymmetric models.
In the context of supergravity models with universal
soft-supersymmetry-breaking and radiative electroweak breaking we
find $R^{\rm susy}_b\lsim0.0004$, which does not shift significantly
the Standard Model prediction ($R^{\rm SM}_b=0.2157$ for $m_t=175\GeV$;
$R^{\rm exp}_b=0.2204\pm0.0020$). We also compute $R_b$ in the minimal
supersymmetric standard model (MSSM), and delineate the region of parameter
space which yields interestingly large values of $R_b$. This region entails
light charginos and top-squarks, but is {\em strongly} restricted by the {\em
combined} constraints from $B(b\to s\gamma)$ and a not-too-large invisible
top-quark branching ratio: only a few percent of the points with $R^{\rm
susy}_b>0.0020\,(1\sigma)$ are allowed.
\end{abstract}

\vspace{1cm}
\begin{flushleft}
\baselineskip=12pt
{CTP-TAMU-25/95}\\
{ACT-10/95}\\
June 1995
\end{flushleft}
\newpage

\setcounter{page}{1}
\pagestyle{plain}
\baselineskip=14pt

\section{Introduction}
Precision tests of the electroweak interactions at LEP have provided the
most sensitive checks of the Standard Model of particle physics. The pattern
that has emerged is that of consistent agreement with the Standard Model
predictions. This pattern seems to have so far only one apparently dissonant
note, namely in the measurement of the ratio $R_b\equiv \Gamma(Z\to b{\bar
b})/\Gamma(Z\to{\rm hadrons})$, where the latest global fit to the LEP data
($R^{\rm exp}_b=0.2204\pm0.0020$ \cite{Rbexp}) lies more than two standard
deviations above the one-loop Standard Model prediction \cite{RbSM} for all
preferred values of the top-quark mass (\eg, $R_b^{\rm SM}=0.2157$ for
$m_t=175\GeV$). Further experimental statistics will reveal whether this is
indeed a breakdown of the Standard Model. In the meantime, it is important to
explore what new contributions to $R_b$ are expected in models of new physics,
such as supersymmetry.

The study of supersymmetric contributions to $\Gamma(Z\to b\bar b)$ has
proceeded in two phases. Originally the quantity $\epsilon_b$ \cite{ABC-Rb} was
defined as an extension of the $\epsilon_{1,2,3}$ scheme \cite{ABC} for
model-independent fits to the electroweak data. More recently it has become
apparent that the ratio $R_b$ is more directly calculable \cite{BF,KKW,GJS} and
readily measurable. It has been made apparent \cite{KKW} that supersymmetric
contributions to $R_b$ are not likely to increase the total predicted value for
$R_b$ in any significant manner, as long as typical assumptions about unified
supergravity models are made. On the other hand, if these assumptions are
relaxed, it is possible for supersymmetry to make a significant contribution to
$R_b$, if certain conditions on the low-energy supersymmetric spectrum are
satisfied~\cite{KKW,GJS}.

In this paper we reexamine this question in the context of a variety of
supergravity models, which include the constrained minimal supergravity model
considered in Ref.~\cite{KKW}, as well as non-minimal string-inspired
supergravity models. In all cases we find $R^{\rm susy}_b\lsim0.0004$, which
brings the Standard Model prediction at most one-fifth of a standard deviation
closer to the experimental result. Moreover, $R^{\rm susy}_b$ could well be
negative, worsening the Standard Model fit, although only slightly.
We then turn to the minimal supersymmetric standard model (MSSM), where the
many parameters are {\em a priori} independent, and seek to delineate the
region of parameter space which yields large values of $R_b$. This exercise
confirms the results of Ref.~\cite{KKW}, that light charginos and top-squarks
of definite composition are required. However, we take the further step of
trying to quantify how large a region of parameter space this is. In other
words, what kind of fine-tuning is involved in attaining such large values of
$R_b$. To this end we apply all known experimental constraints which may affect
the preferred region of parameter space: (i) that $B(b\to s\gamma)$ be within
the allowed CLEO range, (ii) that the lightest supersymmetric particle (LSP) be
neutral and colorless, (iii) that the invisible top-quark branching ratio (\ie,
 $B(t\to\tilde t_1\chi^0_{1,2})$) be within experimental limits, (iv) that the
invisible $Z$ width be within LEP limits, and (v) that
$B(Z\to\chi^0_1\chi^0_2)$ be within LEP limits. By sampling a large number of
random values for the six-plet of parameters that determine $R^{\rm susy}_b$
and the other five experimental observables, we conclude that, of all points in
parameter space that yield $R^{\rm susy}_b>0.0020$ ($1\sigma$), only a few
percent satisfy the {\em combined} additional experimental constraints.

We should remark that the region of MSSM parameter space where $R^{\rm susy}_b$
is enhanced, has recently become the focus of attention for another reason. It
has been argued that the apparent disagreement between the LEP determination of
$\alpha_s$ and the corresponding value obtained from low-energy measurements,
may hint to the possibility of new physics \cite{Shifman}. Moreover, light
supersymmetric particles seem to fit the bill by shifting downwards the
LEP-extracted value of $\alpha_s$ \cite{KSW}, if observables like $R_b$ can be
brought to better agreement with experiment. Our restrictions on parameter
space apply also to this case, to the extent that sufficiently large values of
$R_b$ are called for.

\section{Supersymmetric contributions to $R_b$}
Besides the one-loop Standard Model contributions to $R_b$, in supersymmetric
models there are four new diagrams, as follows:
\begin{itemize}
\item The charged-Higgs--top-quark loop, depends on the charged Higgs mass
and the $t-b-H^\pm$ coupling. For a left-handed $b$ quark the coupling is
$\propto m_{t}/\tan\beta$ whereas for a right-handed $b$ quark it is $\propto
m_{b}\tan\beta$. Therefore, for small\footnote{In supergravity models, the
radiative electroweak breaking mechanism requires $\tan\beta>1$.}  (large)
$\tan\beta$ left- (right)-handed $b$-quark production is dominant. (For
$\tan\beta\gg1$, the value of $m_b$ impacts the contribution significantly.)
It has been shown that the $H^\pm-t$ contribution is always negative
\cite{hollik}, a fact which makes the prediction for $R_b$ in two Higgs-doublet
models always in worse agreement with experiment.
\item The chargino--top-squark loop, is the supersymmetric counterpart of the
$H^\pm$--$t$ loop discussed above. The chargino mass eigenstate is a mixture of
(charged) Higgsino and wino, and the coupling strength is a complicated matter
now because it involves the top-squark mixing matrix and the chargino mixing
matrix. However, because only the Higgsino admixture in the chargino eigenstate
has a Yukawa coupling to the $t$--$b$ doublet, generally speaking a light
chargino with a significant Higgsino component, and a light top-squark
with a significant right-handed component are required for this diagram to make
a non-negligible contribution to $R_b$ \cite{KKW}.
\item The neutralino--bottom-squark loop, is the supersymmetric counterpart of
the neutral-Higgs--bottom-quark loop. The coupling strength of
$\chi_1^{0}-{\tilde b}-b$ is also rather complicated, since it involves the
bottom-squark mixing, the neutralino mixing, and their masses. However, it can
be non-negligible since it is proportional to $m_b\tan\beta$ for the
left-handed $b$-quark. Therefore in the high-$\tan\beta$ region we have to
include this contribution.\footnote{This term was not included in our previous
study in terms of the parameter $\epsilon_b$ \cite{eps1-epsb}, although it was
pointed out that its effects were non-negligible for $\tan\beta\gg1$.}
\item The neutral-Higgs--bottom-quark loop, involves the three neutral scalars,
two CP-even ($h, H$) and one CP-odd ($A$). For the $h\,(H)$ neutral Higgs
boson the coupling to the bottom quark is $\propto m_b \sin\alpha (\cos\alpha)$
which in the absence of a $\tan\beta$ enhancement makes its contribution
negligible. For the $A$ Higgs boson, the coupling to $b\bar b$ is
$\propto\tan\beta$ and the $A$-dependent contribution can be large and positive
if $m_A\lsim90\GeV$ and $\tan\beta\gsim30$ \cite{hollik}. Since we restrict
ourselves to $\tan\beta\lsim20$, and $m_A\gsim100\GeV$ in supergravity models,
this contribution is neglected in what follows.
\end{itemize}
Our computations of $R_b^{\rm susy}$ have been performed using the expressions
given in Ref.~\cite{KKW}. Even though these formulas are given explicitly, the
details are quite complicated by the presence of various Passarino-Veltman
loop functions. As a check of our computations, we have verified numerically
that the results are independent of the unphysical renormalization scale that
appears in the formulas.

\section{$R_b$ in supergravity models}
We consider unified supergravity models with universal soft supersymmetry
breaking at the unification scale, and radiative electroweak symmetry breaking
(enforced using the one-loop effective potential) at the weak scale
\cite{reviews}. These constraints reduce the number of parameters needed to
describe the models to four, which can be taken to be $m_{1/2},
\xi_0\equiv m_0/m_{1/2},\xi_A\equiv A/m_{1/2},\tan\beta$, with a specified
value for the top-quark mass, which we take to be $m_t=175\GeV$. Among
these four-parameter supersymmetric models, we consider generic models
with continuous values of $m_{\chi^\pm_1}$ and discrete choices for the other
three parameters:
\begin{equation}
\tan\beta=2,10,20\ ;\qquad \xi_0=0,1,2,5\ ;\qquad \xi_A=0\ .
\label{generic}
\end{equation}
The choices of $\tan\beta$ are representative; higher values of $\tan\beta$ are
likely to yield values of $B(b\to s\gamma)$ in conflict with present
experimental limits \cite{bsgamma}. The choices of $\xi_0$ correspond to
$m_{\tilde q}\approx(0.8,0.9,1.1,1.9)m_{\tilde g}$. The choice of $A$ has
little impact on the results.
We also consider the case of no-scale $SU(5)\times U(1)$ supergravity
\cite{reviews}. In this class of models the supersymmetry breaking parameters
are related in a string-inspired way. In the two-parameter {\em moduli}
scenario $\xi_0=\xi_A=0$ \cite{LNZI}, whereas in the {\em dilaton} scenario
$\xi_0={1\over\sqrt{3}},\ \xi_A=-1$ \cite{LNZII}.
A series of experimental constraints and predictions for these models have been
given in Ref.~\cite{Easpects}. In particular, the issue of precision
electroweak tests in this class of models has been addressed in
Refs.~\cite{ewcorr,eps1-epsb}.

\begin{figure}[p]
\vspace{6in}
\includegraphics{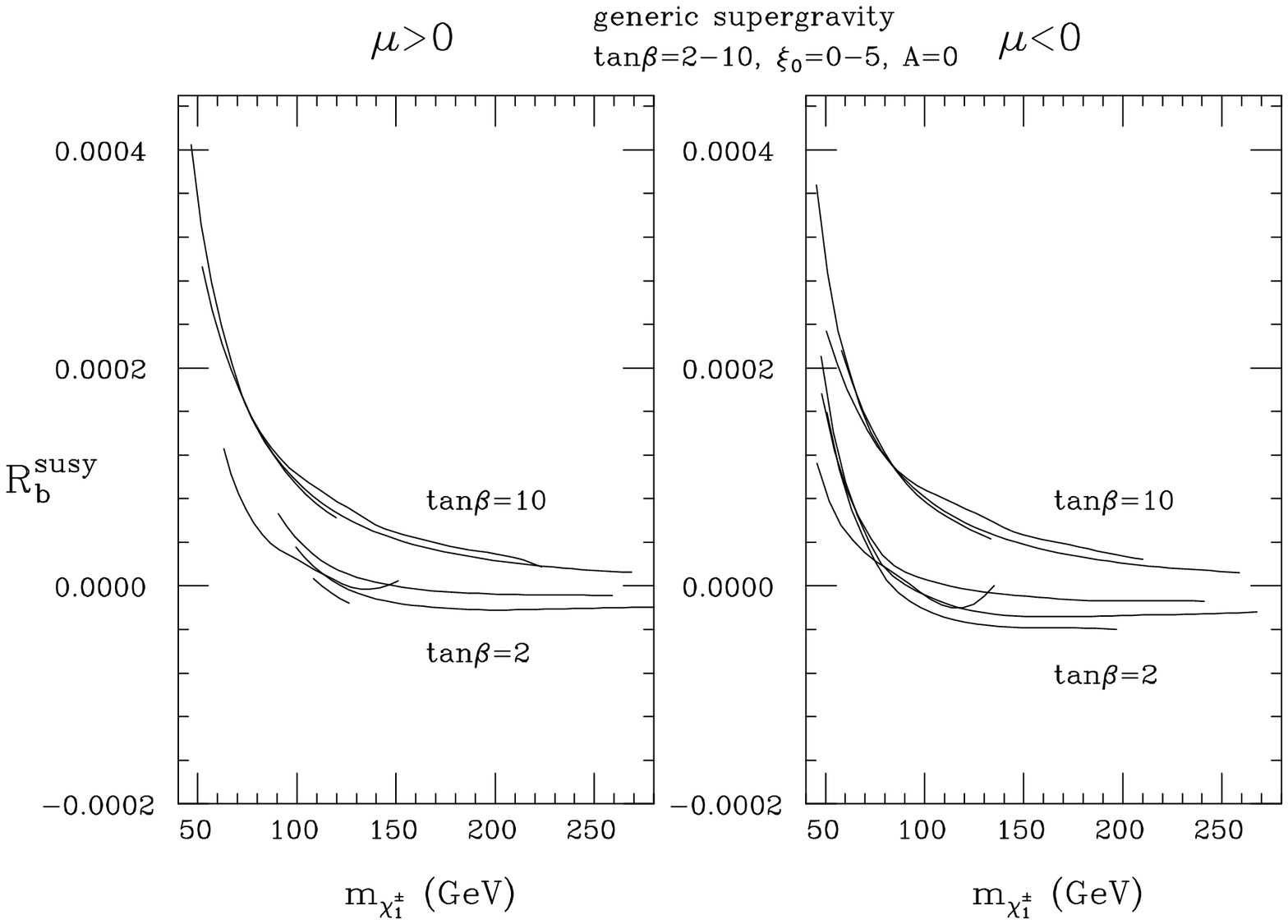}
\caption{The supersymmetric contribution to $R_b$ as a function of the chargino
mass in generic supergravity models with $\tan\beta=2,10$, $\xi_0=0-5$, and
$A=0$. Curves for other values of $\tan\beta$ fall between the two sets shown.}
\label{Rb_SSM}
\end{figure}

\begin{figure}[p]
\vspace{6in}
\includegraphics{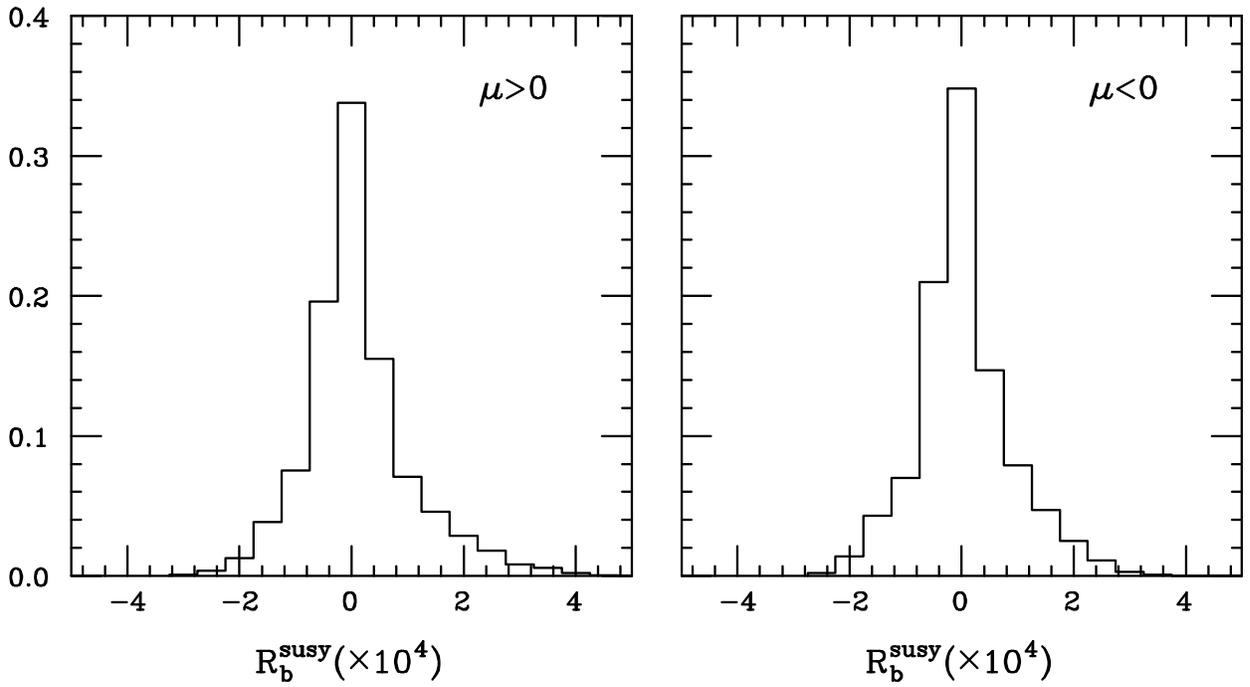}
\caption{Relative distribution of calculated values of $R^{\rm susy}_b$ in a
sample of 10,000 random choices for the four-plet of supergravity parameters.}
\label{hist.SSM}
\end{figure}

The predictions for $R^{\rm susy}_b$ in the generic supergravity models are
shown in Fig.~\ref{Rb_SSM}. Only curves for $\tan\beta=2,10$ are shown, since
the corresponding sets of curves for other values of $\tan\beta$ fall between
these two sets of curves. The results in $SU(5)\times U(1)$ supergravity, for
the same values of the parameters, differ little from those in the generic
models. To add generality to our result and to compare with the study made
below in the case of the MSSM, we have also considered the case where the four
supergravity parameters are chosen at random. We have sampled 10,000 random
four-plets of parameter values, in the ranges: $50\le m_{1/2}\le500\GeV$,
$0\le\xi_0\le5$, $-5\le\xi_A\le5$, and $1\le\tan\beta\le40$. As expected, we
find $R^{\rm susy}_b\lsim0.0004$. A histogram of the relative distribution of
$R^{\rm susy}_b$ values is shown in Fig.~\ref{hist.SSM}, which shows that
$R^{\rm susy}_b$ is equally likely to be positive or negative, and that the
``preferred value" is very small.

\begin{figure}[p]
\vspace{6in}
\includegraphics{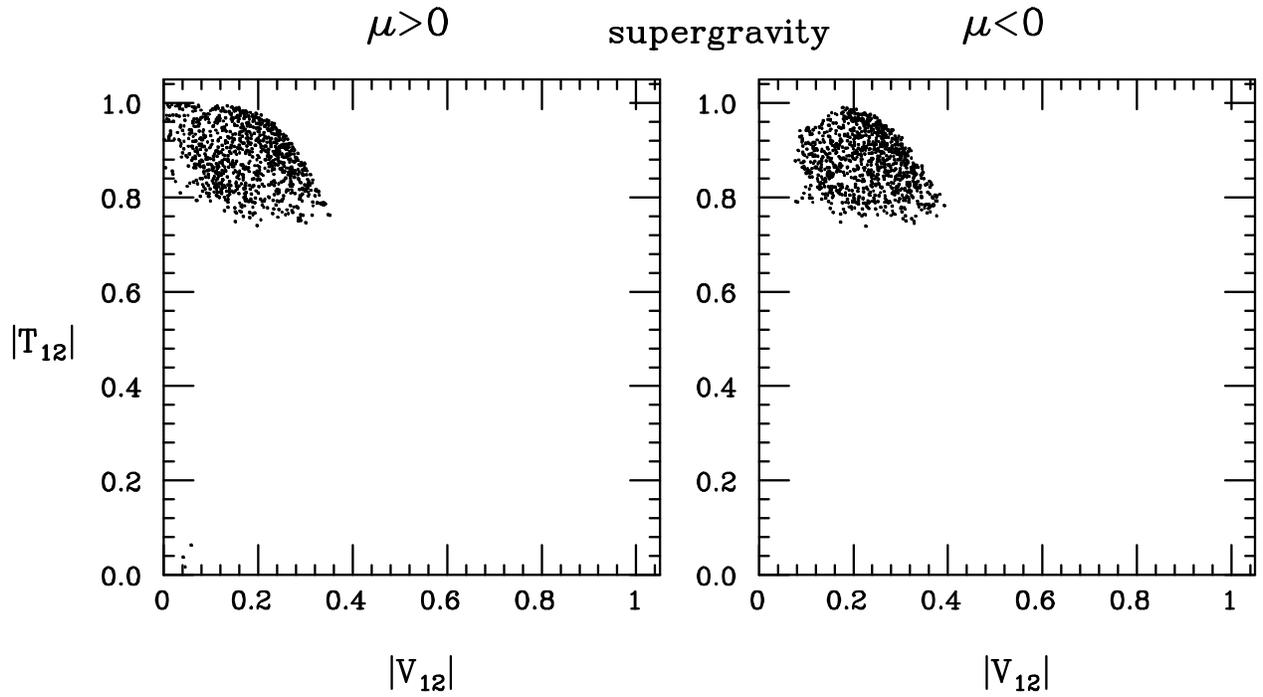}
\caption{The right-handed component of the lightest top-squark ($T_{12}$)
versus the higgsino component of the lightest chargino ($V_{12}$) in a
random sample of supergravity models. Large values of both $T_{12}$
and $V_{12}$ are required for an enhanced value of $R^{\rm susy}_b$. }
\label{TvsV.SSM}
\end{figure}

In almost all cases the largest positive contribution to $R_b^{\rm susy}$ comes
from the chargino--top-squark loop. As expected, the largest contribution
from this diagram happens for points with the lightest chargino masses (which
correspond to the lightest ${\tilde t}_1$ masses) since supersymmetry is a
decoupling theory. However, even the largest value $(\approx0.0004)$ is still
very small compared with the largest possible result in a generic low-energy
supersymmetric model. The reason for this is that while the smallest possible
chargino and a top-squark masses are required for an enhanced contribution,  it
is also necessary that the top-squark be mostly right-handed and that the
chargino has a significant Higgsino component~\cite{KKW}.  A scatter plot of
the values of these couplings is shown in Fig.~\ref{TvsV.SSM}, where the
right-handed component of the lightest top-squark ($T_{12}$) is plotted against
the higgsino component of the lightest chargino ($V_{12}$). The first
requirement is in fact attainable in these models (\ie, $|T_{12}|\sim1$), but
the former is not (\ie, $|V_{12}|\lsim0.4$). This behavior is expected in
supergravity models with radiative electroweak symmetry breaking since for
light charginos $|\mu|\gg M_2$, which makes the lightest chargino mainly a wino
instead of a Higgsino. (This is also the reason why the results in $SU(5)\times
U(1)$ supergravity differ little from those in the generic models.) Concerning
the other contributions to $R^{\rm susy}_b$, the charged-Higgs--top-quark loop
is always negative, and is enhanced for either small or large values of
$\tan\beta$. The neutralino--bottom-squark contribution is almost always
smaller than the chargino--top-squark contribution and not of definite sign.

Thus, $R_b$ in supergravity models could improve the Standard Model
fit to the LEP data by at most one-fifth of a standard deviation, and it could
well worsen the fit by the same small amount.

\section{$R_b$ in the MSSM}
We now relax the supergravity assumptions that correlate the various
supersymmetric parameters, in order to explore the region of the MSSM
parameter space that yields large values of $R_b$. This region is then
subjected to all available experimental constraints, which have the effect of
restricting it significantly. The $R^{\rm susy}_b$ observable depends
on six basic parameters: the elements of the chargino mass matrix
($M_2,\mu,\tan\beta$) which determine the chargino masses and their
composition, and the two top-squark ($\tilde t_1,\tilde t_2$) masses and their
mixing angle ($\theta_{\tilde t}$). In addition, the Higgs boson masses
enter, although for $\tan\beta<30$ the neutral Higgs boson contribution is
not relevant \cite{KKW}. The charged Higgs mass is relevant for small values
of $\tan\beta$, but this contribution to $R_b$ is always negative and will
be neglected in what follows (\ie, we take a large charged Higgs mass).
Neglecting altogether the Higgs-boson contribution to $R^{\rm susy}_b$ is
generally justified when looking for the largest values of $R_b$, although
some exceptions exist for rather low values of the pseudoscalar Higgs boson
mass ($m_A$) and large values of $\tan\beta$ \cite{GJS}.

\begin{figure}[p]
\vspace{6in}
\includegraphics{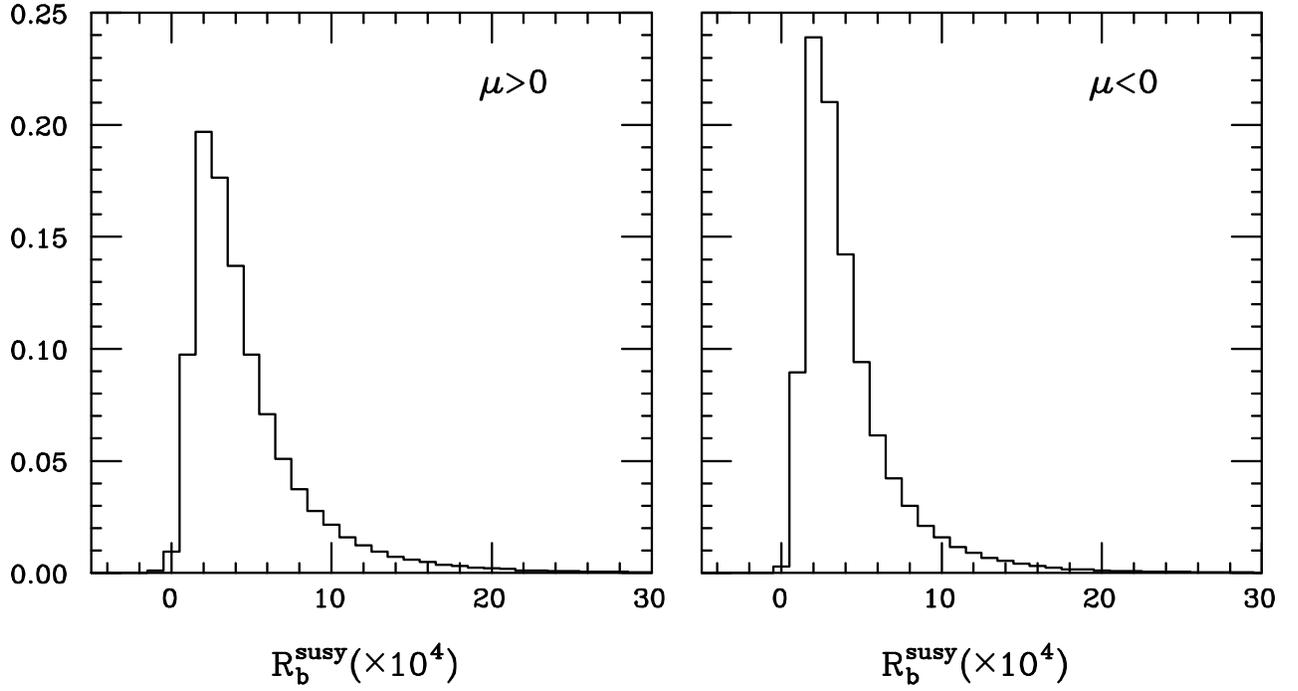}
\caption{Relative distribution of calculated values of $R^{\rm susy}_b$ in a
large sample of random choices for the six-plet of relevant MSSM parameters.
For $\mu\!\!>\!\!0\,(\mu\!\!<\!\!0)$, only $\approx0.9\%\,(0.5\%)$ of the
points yield $R^{\rm susy}_b>0.0020\,(1\sigma)$.}
\label{hist.MSSM}
\end{figure}

We have sampled a large number of random choices for the six-plet of
parameters, in the ranges $0\le M_2,|\mu|,m_{\tilde t_1},m_{\tilde
t_2}\le250\GeV$, $1\le\tan\beta\le5$, and $0\le\theta_{\tilde t}\le\pi$. We
have concentrated on the small-$\tan\beta$ region since for large values of
$\tan\beta$, where $R^{\rm susy}_b$ may also be enhanced, $B(b\to s\gamma)$ is
likely to exceed significantly the allowed CLEO range \cite{bsgamma}. A
histogram depicting the relative distribution of calculated values of $R^{\rm
susy}_b$ is shown in Fig.~\ref{hist.MSSM}. [Note that small negative values of
$R^{\rm susy}_b$ are possible (although not very likely).] As expected, there
is a steady decline in the likelihood of the larger values of $R^{\rm susy}_b$,
with the following relative distributions
\begin{equation}
\begin{tabular}{ccrr}
Condition&Comment&$\mu>0$&$\mu<0$\\ \hline
$R^{\rm susy}_b>0.0008$&Ok at 95\%CL&12\%&8.8\%\\
$R^{\rm susy}_b>0.0014$&Ok at 90\%CL&3.0\%&1.9\%\\
$R^{\rm susy}_b>0.0020$&$R^{\rm SM}_b$ up by $1\sigma$&0.87\%&0.52\%\\
$R^{\rm susy}_b>0.0027$&Ok at $1\sigma$&0.20\%&0.12\%
\end{tabular}
\label{distributions}
\end{equation}
where for example, only $\sim(2-3)\%$ of the sampled six-plets yield $R^{\rm
susy}_b>0.0014$, which brings the total $R_b$ prediction inside the 90\%CL
allowed range for $m_t=175\GeV$. Also, to be within the $1\sigma$ range, and
thus to significantly improve the fit to the LEP data, requires considerable
fine tuning. If the sampling region were to be extended (\ie, a larger mass
interval or larger values of $\tan\beta$) we would find that the new points
fall into the lowest bins in Fig.~\ref{hist.MSSM}, while the population of the
bins with large $R_b$ values would correspondingly decrease. The distributions
in Eq.~(\ref{distributions}) would change accordingly.

\begin{figure}[p]
\vspace{6in}
\includegraphics{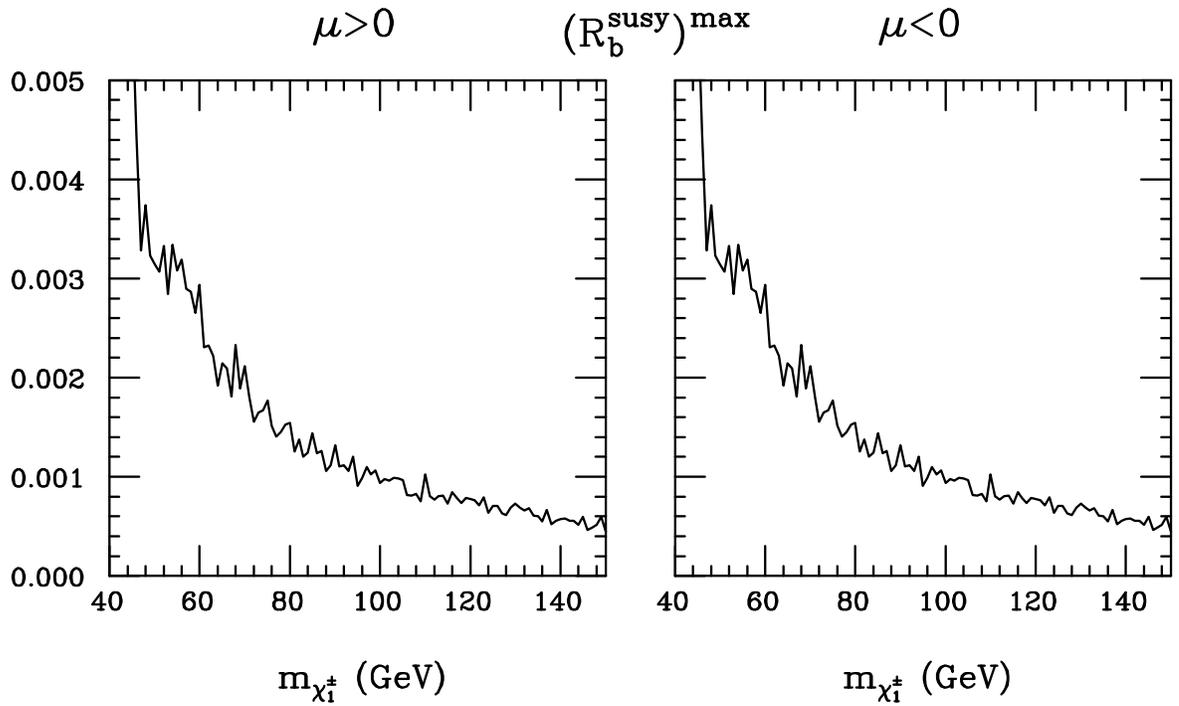}
\caption{Maximum attainable value of $R^{\rm susy}_b$ in the MSSM as a function
of the lightest chargino mass. All other variables have been integrated out.
The broken line is a figment of the finite sample size.}
\label{Rbmax}
\end{figure}

In Fig.~\ref{Rbmax} we show the maximum attainable value of $R^{\rm susy}_b$ as
a function of the chargino mass, which makes apparent the need for a light
chargino if $R^{\rm susy}_b$ is to be enhanced. In fact, the maximum value is
obtained for $m_{\chi^\pm_1}\to {1\over2}M_Z$ (and also
$m_{\tilde t_1}\to{1\over2}M_Z$).
This phenomenon has been observed before in the context of the $\epsilon_b$
approach to the problem \cite{ABC-Rb,ewcorr}, and corresponds to a wavefunction
renormalization effect when the particles in the loop go on shell.

\begin{figure}[p]
\vspace{6in}
\includegraphics{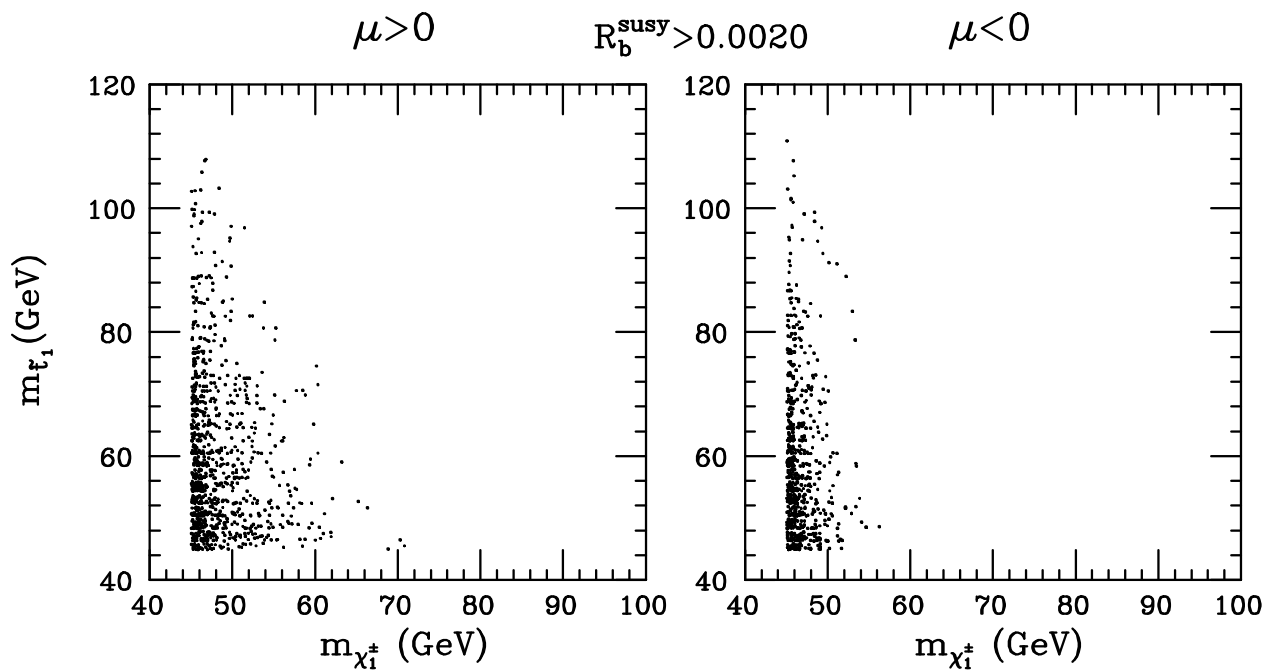}
\caption{The distribution of lightest chargino and top-squark masses for
points in MSSM parameter space with $R^{\rm susy}_b>0.0020\,(1\sigma)$.}
\label{stop-ch}
\end{figure}

\begin{figure}[p]
\vspace{6in}
\includegraphics{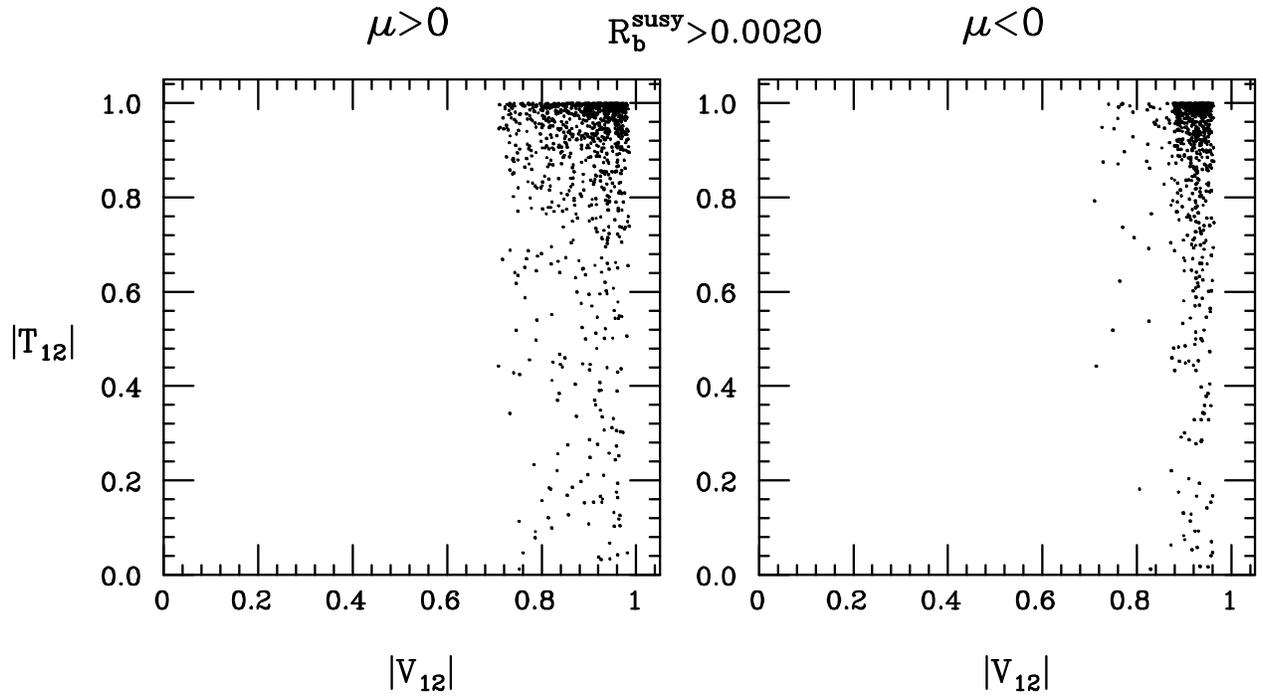}
\caption{The right-handed component of the lightest top-squark ($T_{12}$)
versus the higgsino component of the lightest chargino ($V_{12}$) for the
fraction of MSSM sampled points with $R^{\rm susy}_b>0.0020\,(1\sigma)$.}
\label{TvsV.MSSM}
\end{figure}

For concreteness, in what follows we concentrate on the $\lsim1\%$ of the
points that increase $R_b$ by at least $1\sigma$; the other cases in
Eq.~(\ref{distributions}) depend on the choice of $m_t$. By design, this sample
contains 1000 points for each sign of $\mu$. First let us show that the region
of parameter space which leads to enhanced values of $R^{\rm susy}_b$ is indeed
characterized by light higgsino-like charginos and light right-handed--like
top-squarks. In Fig.~\ref{stop-ch} we show the distribution of points in the
$(m_{\chi^\pm_1},m_{\tilde t_1})$ plane that lead to $R^{\rm susy}_b>0.0020$,
and in Fig.~\ref{TvsV.MSSM} we show the corresponding distribution in
$(|V_{12}|,|T_{12}|)$ space. If $R^{\rm susy}_b>0.0020$ is indeed required,
then LEPII should see the lightest chargino and possibly also the ligthest
top-squark. The observed marked preference for large values of the
$V_{12},T_{12}$ admixtures becomes evident when one considers the top-quark
Yukawa coupling
\begin{equation}
\lambda_t\, Q_3 \widetilde H \widetilde t_R\to \lambda_t\, b_L \widetilde
H^\pm\widetilde t_R\,,\ \lambda_t\, t_L \widetilde H^0 \widetilde t_R\ ,
\label{Yukawa}
\end{equation}
which picks out $\widetilde t_1\to\widetilde t_R$, $\chi^\pm_1\to\widetilde
H^\pm$, $\chi^0_1\to\widetilde H^0$. This interaction also leads to an enhanced
coupling between the top-quark, a higgsino-like neutralino, and a right-handed
top-squark, and may lead to enhanced exotic decays of the top-quark, as we
discuss below. Comparing Fig.~\ref{TvsV.SSM} with Fig.~\ref{TvsV.MSSM} we see
why $R^{\rm susy}_b$ is always suppressed in supergravity models: the regions
in ($|V_{12}|,|T_{12}|$) space are completely non-overlapping. We should note
that for $\mu>0$ it is easier to obtain large-$R^{\rm susy}_b$ solutions since
the chargino can have a larger higgsino component (as Fig.~\ref{TvsV.MSSM}
shows), \ie, $|V_{12}|=|\cos\phi_+|$ with
$\tan(2\phi_+)\propto-\mu\cos\beta+M_2\sin\beta$ \cite{HK}.

\begin{table}[t]
\caption{The 95\%CL upper limit on the invisible top-quark branching ratio,
obtained from a comparison of D0 data with theoretical expectations. All
masses in GeV, all cross sections in pb.}
\label{Table1}
\begin{center}
\begin{tabular}{ccccc}
$m_t$&$\sigma^{\rm D0}_{\rm min}$&$\sigma^{\rm th}_{\rm max}$
&$B^{t\to bW}_{\rm min}$&$B^{t\to{\rm inv}}_{\rm max}$\\ \hline
150&3.6&13.8&0.51&0.49\\
160&3.4&9.53&0.60&0.40\\
170&3.1&6.68&0.68&0.32\\
180&2.5&4.78&0.72&0.28\\
190&2.2&3.44&0.80&0.20\\
200&2.1&2.52&0.91&0.09
\end{tabular}
\end{center}
\hrule
\end{table}

The parameters that determine $R^{\rm susy}_b$ also determine the following
observables:
\begin{enumerate}
\item $B(b\to s\gamma)$, as measured by CLEO to be $(1-4)\times10^{-4}$
\cite{CLEO}, has been computed as in Ref.~\cite{bsgamma}. The charged-Higgs
loop is negligible in the limit of a large charged Higgs mass (as assumed
above) and in general  its contribution is not large and does not
significantly affect whether $B(b\to s\gamma)$ is in agreement with experiment
or not.
\item $m_{\chi^0_1}<\{m_{\chi^\pm_1},m_{\tilde t_1}\}$, is the cosmological
requirement of a neutral and colorless lightest supersymmetric particle (LSP).
The lightest neutralino mass follows from the inputs to the chargino mass
matrix and the usual assumption relating $M_1$ to $M_2$. This constraint has
already been included in Eq.~(\ref{distributions}) and in
Figs.~\ref{hist.MSSM},\ref{stop-ch},\ref{TvsV.MSSM}.
\item $B(t\to\tilde t_1\chi^0_{1,2})$ leads to an ``invisible" decay of the
top quark if the $\tilde t_1\to b\chi^\pm_1\to b(q\bar
q',\ell\nu_\ell)\chi^0_1$ decay products do not pass the standard top-quark
cuts, as may likely be the case given the larger amount of missing energy and
the softer leptons that accompany this decay. Taking the 95\% CL lower bound on
the top-quark cross section from D0 (see Fig. 3 in Ref.~\cite{D0}), and
dividing it by the ``upper" estimate of the theoretical cross section (see
Table 1 in Ref.~\cite{LSvN}), one can obtain a 95\%CL lower bound on $[B(t\to
Wb)]^2$. As a function of $m_t$, this exercise is carried out in
Table~\ref{Table1}. For $m_t=175\pm15\GeV$ we obtain
$B^{t\to{\rm inv}}_{\rm max}=0.3\mp0.1$.
\item The invisible $Z$ width $\Gamma(Z\to\chi^0_1\chi^0_1)$ follows
from the neutralino mass and composition.
It should not exceed 7.6 MeV \cite{Zinv}.
\item The branching ratio $B(Z\to\chi^0_1\chi^0_2)$ also follows from the
neutralino mass matrix, and should not exceed $10^{-4}$ \cite{L3}.
\end{enumerate}

\begin{table}[t]
\caption{The fraction of MSSM parameter space with $R^{\rm susy}_b>0.0020$,
where the five experimental constraints are satisfied individually, and the
combined allowed region, for $B^{\rm t\to inv}_{\rm max}=0.3\pm0.1$.}
\label{Table2}
\begin{center}
$\mu>0$\qquad
\begin{tabular}{c|ccccc|c}
$B^{\rm t\to inv}_{\rm max}$&$B^{b\to s\gamma}$&LSP&$B^{\rm t\to
inv}$&$\Gamma^{\rm inv}_Z$&$B^{Z\to\chi^0_1\chi^0_2}$&All\\ \hline
0.4&20\%&98\%&41\%&88\%&72\%&5.3\%\\
0.3&20\%&98\%&19\%&88\%&72\%&1.7\%\\
0.2&20\%&98\%&8.0\%&88\%&72\%&0.9\%
\end{tabular}

\bigskip
$\mu<0$\qquad
\begin{tabular}{c|ccccc|c}
$B^{\rm t\to inv}_{\rm max}$&$B^{b\to s\gamma}$&LSP&$B^{\rm t\to
inv}$&$\Gamma^{\rm inv}_Z$&$B^{Z\to\chi^0_1\chi^0_2}$&All\\ \hline
0.4&37\%&97\%&31\%&98\%&95\%&8.9\%\\
0.3&37\%&97\%&14\%&98\%&95\%&1.9\%\\
0.2&37\%&97\%&8.4\%&98\%&95\%&0.6\%
\end{tabular}
\end{center}
\hrule
\end{table}

\begin{figure}[p]
\vspace{6in}
\includegraphics{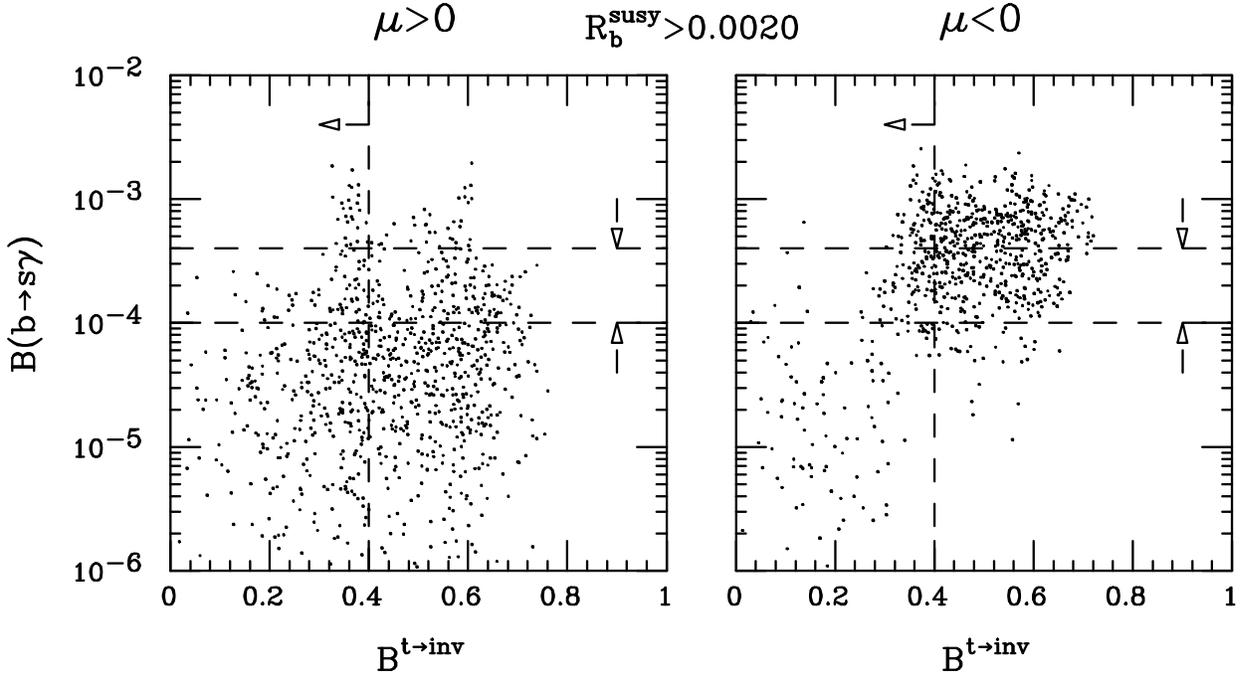}
\caption{The calculated value of $B(b\to s\gamma)$ versus $B^{\rm t\to inv}$
for points in the MSSM parameter space with $R^{\rm susy}_b>0.0020$. The
arrows point into the experimentally allowed region. For
$\mu\!\!>\!\!0\,(\mu\!\!<\!\!0)$ only $\approx5\%\,(9\%)$ of the points fall
inside the combined allowed region.}
\label{bsg-inv}
\end{figure}

\begin{figure}[p]
\vspace{6in}
\includegraphics{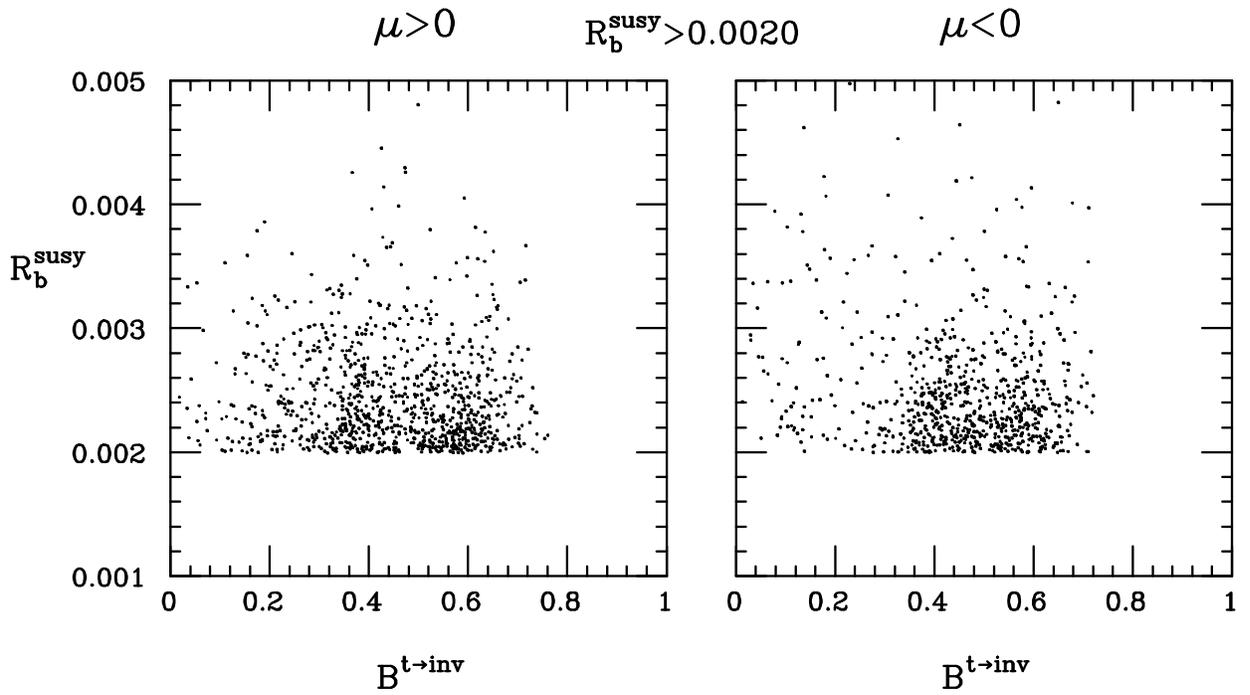}
\caption{The calculated value of $R^{\rm susy}_b$ versus $B^{\rm t\to inv}$
for points in the MSSM parameter space with $R^{\rm susy}_b>0.0020$. Note
the correlated enhancement of these two observables.}
\label{Rb-inv}
\end{figure}

We now impose these additional constraints on the points in MSSM parameter
space with $R^{\rm susy}_b>0.0020$. First let us give the fraction of parameter
space where these constraints are satisfied individually, as well as the
combined allowed region. In Table~\ref{Table2} we carry out this exercise for
$B^{\rm t\to inv}=0.3\pm0.1$, which corresponds to $m_t=175\mp15\GeV$. One can
see that only a few percent of the points with interestingly large values of
$R^{\rm susy}_b$ are experimentally allowed. Table~\ref{Table2} also shows
that the $b\to s\gamma$ and $B^{\rm t\to inv}$ constraints are the most
restricting ones, and that they are both satisfied only in a small region of
parameter space. This trend is shown more concretely in Fig.~\ref{bsg-inv},
where the distribution in $[B^{\rm t\to inv},B(b\to s\gamma)]$ space is shown,
together with the present experimental limits on these observables. In this
figure, for $\mu>0\,(\mu<0)$: $\sim20\,(37)\%$ of the points fall inside the
allowed $B(b\to s\gamma)$ region, $\sim41\,(31)\%$ of the points fall inside
the allowed  $B^{\rm t\to inv}$ area, and $\sim5\,(9)\%$ fall in the combined
allowed area. As remarked above, in regions of parameter space with enhanced
values of $R^{\rm susy}_b$, we may also expect enhanced values of $B^{\rm t\to
inv}$. In Fig.~\ref{Rb-inv} we show the distribution of $R^{\rm susy}_b$ versus
$B^{\rm t\to inv}$  calculated values (with a cutoff of $R^{\rm
susy}_b>0.0020$), which indeed shows a distinct correlated enhancement of these
two observables.

\begin{figure}[p]
\vspace{6in}
\includegraphics{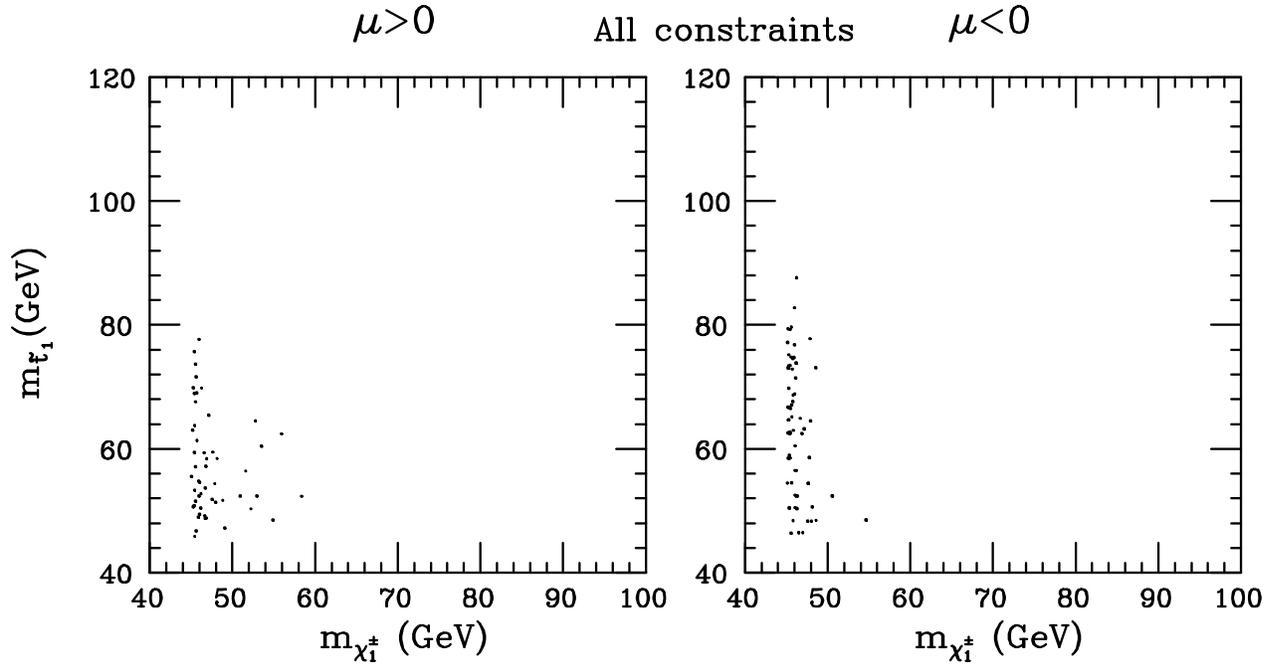}
\caption{The distribution of lightest chargino and top-squark masses for
points in MSSM parameter space with $R^{\rm susy}_b>0.0020\,(1\sigma)$, when
{\em all} experimental constraints are imposed (with $B^{\rm t\to inv}_{\rm
max}=0.4$). Because of the finite sample size, the whole region spanned by the
shown points should be considered viable.}
\label{stop-ch.OkOk}
\end{figure}

Finally, we consider the case where {\em all} experimental constraints are
applied simultaneously. The resulting allowed points in the
$(m_{\chi^\pm_1},m_{\tilde t_1})$ plane, for $B^{\rm t\to inv}_{\rm max}=0.4$,
are shown in Fig.~\ref{stop-ch.OkOk}. This figure is to be contrasted with
Fig.~\ref{stop-ch}, where none of the experimental constraints were applied
(except for the rather unrestricting LSP constraint). We see that the extent
of the allowed area is reduced: $m^{\rm max}_{\chi^\pm_1}:(70\to60)\GeV$ and
$m^{\rm max}_{\tilde t_1}:(110\to80)\GeV$. This restricted parameter space
should be even more straightforwardly tested at LEPII, with now guaranteed
sensitivity to the ligthest top-squark. If we strengthen the $B^{\rm t\to inv}$
constraint to $B^{\rm t\to inv}_{\rm max}=0.2$, the more relevant
shrinking of parameter space occurs in the $m_{\tilde t_2}$ and $\tan\beta$
directions: $m_{\tilde t_2}<100\GeV$ and $\tan\beta\lsim1.5$ are now required.
Such strong restrictions would also allow detection of the lightest Higgs boson
at LEPII, and both top-squarks at the Tevatron and LEPII.

\section{Conclusions}
We have studied the contributions to $R_b$ that are expected in low-energy
supersymmetric models. In the case of supergravity models with radiative
electroweak symmetry breaking, $R^{\rm susy}_b$ could improve the
Standard Model fit to the LEP data by only a small fraction of a standard
deviation, and it could well worsen the fit by the same small amount. In the
MSSM, light higgsino-like charginos and light right-handed-like top-squarks
are required to obtain a sizeable $R^{\rm susy}_b$ value. Such values occur in
a small fraction of the parameter space, which is largely accessible to LEPII
searches. This region of parameter space is further constrained by five
experimental observables, most importantly $B(b\to s\gamma)$ and the
invisible top-quark branching fraction. Imposing these additional constraints
reduces the allowed parameter space to a few percent of its unconstrained
size, and makes experimental exploration of this scenario in both
chargino and top-squarks assured at LEPII.

\vspace{1cm}
\section*{Acknowledgements}
This work has been supported in part by DOE grant DE-FG05-91-ER-40633. The
work of X. W. has been supported by the World Laboratory. We would like to
thank Chris Kolda for helpful communications. X. W. would like to thank
J.~T.~Liu for helpful discussions. J. L. would like to thank James White
and Teruki Kamon for helpful discussions.

\clearpage

\clearpage

\clearpage

\clearpage

\clearpage

\clearpage

\clearpage

\clearpage

\clearpage

\clearpage

\end{document}